# An expression for the eddy field in a circular vacuum chamber for HEPS booster dipole

Y.Chen[a,*], W.Kang[a], Y.M.Peng[a], J.Y.Li[a], D.Wang[a], C.H.Yu[a], S.Wang[a], Q.Qin[a]

[a] Institute of High Energy Physics, CAS, Yuquan Road 19B, Beijing 100049, China

**Abstract:** The analytical expression of the magnetic field distribution within the aperture of a circular vacuum chamber due to the induced eddy is derived. Two cases are discussed, one is the absence of iron, the other is that the vacuum chamber is between the iron poles, that implies the use of the image current methods. The current angular distribution in the vacuum chamber can be calculated from the ramping rate of the exposed field, then the contour integration is applied to the circular current to obtain the field distribution. The formula can be used to estimate the undesired fields from a circular beam box when it is exposed to a ramping field.

**Key words:** Eddy current; Ramping field; Contour integration; Multipole analysis; Magnetic field Simulation

## 1 Introduction

The High Energy Photon Source (HEPS), a 6 GeV green-field diffraction-limited storage ring light source, will be built in Huairou District, Beijing, China. The HEPS design is designed to generate and deliver X-ray synchrotron radiation with high brightness of $5 \times 10^{22}$ photons s$^{-1}$ mm$^{-2}$ mrad$^{-2}$ (0.1% bandwidth)$^{-1}$. The CDR is now finished and ready for construction [1]. The HEPS consists of the linac, booster and the storage ring, the linac provides 300 MeV electron beams, in booster the beam will be accelerate to 6 GeV further, and then injected into the storage ring. The booster is an electron synchrotron with a circumference of 454m, employs 128 dipoles, 148 quadrupoles, 68 sextupoles and 88 correctors. Magnets except correctors are operated under ramping exciting, the cycling frequency is 1$Hz$. The magnetic field will rise to 0.71T from 0.035T within the 0.4s, the ramping speed is 1.6875T/s. Due to this, the eddy current will induced in the circular vacuum chamber. The sextupole magnetic field from the eddy current has long been established, which can be expressed as [2]

$$B_y = \frac{\mu_0 F h}{\rho g} \frac{dB}{dt} x^2 \tag{1}$$

---

* Corresponding author. Tel.: +8610 88236235; fax: +8610 88236190.

E-mail address: chenyuan@ihep.ac.cn (Yuan.Chen).

where $\rho$ is the electrical resistivity, *h* and *g* are respectively the chamber wall thickness and chamber height and *dB/dt* is the ramping rate of the magnetic field of the booster dipole magnet. *F* is a factor determined by the chamber geometry. This formula is based on the parallel top and bottom of the vacuum chamber, for a circular chamber, that will introduce some errors. This motivates us to look for the more precise field distribution within the aperture of chamber. In Section 2, the case of the absence of iron is considered. Section 3 and section 4 present the process of how to take iron into account and the application on HEPS booster dipole respectively. In Section 5, the chromaticity is discussed and a concise summery is given in Section 6.

## 2. Vacuum chamber in iron-free space

When a dipole is AC excited, the eddy current in the yoke will be induced, at the same time the circular metal vacuum chamber is also exposed to a rapidly varying magnetic field, the eddy current will be induced in the conductor, Fig.1 shows 2D situation, *R* is the inner radius of the chamber and the meaning of *t* is the same as Sec.1.

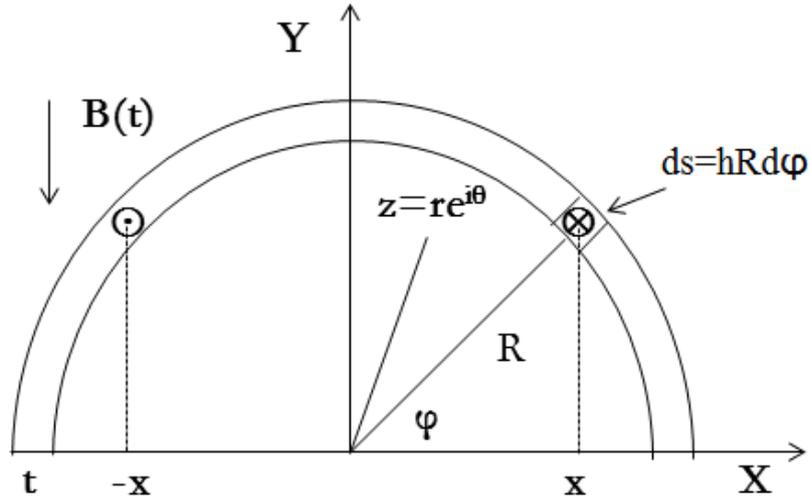

Fig.1. Eddy current induced in the circular conductor vacuum chamber (Half)

Consider a conductor strip along the z-axis with the length *L*, the eddy current will flow into paper at $x=R\cos(\theta)$ and return at $-x$, the varying rate of magnetic flux across the surface enclosed by strips and the resistance of the loop $R_e$ is respectively:

$$-\frac{d\Phi}{dt} = -2RLB'\cos\varphi, \quad B' = \frac{dB}{dt} \tag{2}$$

$$R_e = \rho\frac{L}{s} = \frac{2\rho L}{hRd\varphi}, (L \gg R) \tag{3}$$

Since the only 2D case is under considered, in formula 3, the contribution to the loop resistance from the transversal segments is omitted. The angular distribution of the eddy current in the vacuum chamber has the form

$$I(\varphi) = -\frac{d\Phi/dt}{R_e} = -\frac{hB'R^2\cos\varphi}{\rho}d\varphi \tag{4}$$

where ρ is the resistivity of the chamber. If the magnetic field is a pure dipole field, from formula (4), the angular distribution of the eddy current has a standard cosine form. The magnetic field $B(z)$ due to an infinite current filament located at $Re^{i\varphi}$ is given by [3]

$$B(z = re^{i\theta}) = B_y + iB_x = \frac{\mu_0 I}{2\pi} \frac{1}{z - Re^{i\varphi}} \qquad (5)$$

where z is the measured location and $|z|<R$, i is the complex unit with $i^2=-1$. Inserting the $I(\theta)$ and integrating (5) along the contour $Re^{i\varphi}$, the field in the chamber aperture is

$$B_x = 0$$
$$B_y = \frac{\mu_0 hRB'}{2\rho} \qquad (6)$$

It is clear that a pure and ramping dipole field will only induce a pure dipole in the aperture of the circular conductor chamber due to the eddy current.

## 3 Vacuum chamber between iron poles

When the vacuum chamber is between iron poles, a pure dipole field will induce the same eddy angular distribution in the chamber conductor. Assume the permeability of two poles is infinity, the source current filaments in gap will have infinite image current filaments with the same current. A single current filament located at $Re^{i\varphi}$ between two parallel iron plates has the form [4]

$$B(z) = \frac{\mu_0 I(\varphi)}{4g}\left[\tanh\frac{\pi(z-Re^{-i\varphi})}{2g} + \coth\frac{\pi(z-Re^{i\varphi})}{2g}\right] \qquad (7)$$

where g is the dipole gap. Substituting (4) into (7) and performing the contour integral with some transformations

$$\zeta = Re^{i\varphi}, \cos\varphi = \frac{\zeta + R^2\zeta^{-1}}{2R}, d\varphi = \frac{d\zeta}{i\zeta} \qquad (8)$$

The $B(z)$ in the aperture can be written as

$$B(z) = \frac{\mu_0 hB'R^2}{4g\rho}\int_0^{2\pi}\left[\tanh\frac{\pi(z-\zeta^*)}{2g} + \coth\frac{\pi(z-\zeta)}{2g}\right]\left(\frac{1}{2iR} + \frac{R}{2i\zeta^2}\right)d\zeta \qquad (9)$$

where $\zeta^*$ is the complex conjugate of $\zeta$. The first integral of tanh has one singular pole of order 2 at $\zeta=0$, the second integral of coth has two isolated poles, one at $\zeta=0$ of order 2, the other at $\zeta=z$ of order 1, since the contour $|\zeta|=R$ contains z. By the Residue Theorem, $B(z)$ can be integrated

$$B(z) = \frac{\mu_0 hB'R^2}{4g\rho}\left[\frac{\pi^2 R}{2g}\left(\tanh^2(\frac{\pi z}{2g}) + \coth^2(\frac{\pi z}{2g}) - 2\right) - \left(\frac{2g}{R} + \frac{2gR}{z^2}\right)\right] \qquad (10)$$

With the series expansion of *tanh(z)* and *coth(z)*,

$$\tanh(z) = z - \frac{z^3}{3} + \cdots$$

$$\coth(z) = \frac{1}{z} + \frac{z}{3} - \frac{z^3}{45} + \cdots \tag{11}$$

And let z locate on the x-axis, i.e. $y=0$, keep the series up to order 2, (10) can be expanded as

$$B_y(x, y = 0) = \frac{\mu_0 h B'R^2}{4g\rho}\left(\frac{2g}{R} + \frac{2\pi^2 R}{3g} - \frac{2}{15}\frac{\pi^4 R}{g^3}x^2 + \cdots\right) \tag{12}$$

It is interesting that the divergence term of $z^{-2}$ in (10) cancels with the divergence term of the series of $\coth^2(z)$, that avoids the field divergence at $z=0$. Also, the (12) shows the series of $B(z)$ only includes even order terms of $z$ as it should, since the vacuum chamber is symmetric with the vertical mid-plane.

## 4. Applied to HEPS booster dipole

The booster of HEPS employs H-type laminated dipoles, the gap is 34mm, and center field rises from $0.035T$ to $0.71T$ within $0.4s$, the cross section of the dipole is shown in Fig.2. The field uniformity reaches $5\times10^{-5}$ within $x=\pm15mm$, so the field penetrating the chamber is a pure dipole approximately. The vacuum chamber is made of 304# stainless steel, the resistivity is $73\times10^{-8}\Omega m$, and wall thickness is $0.7mm$, the inner radius is $15mm$.

According to the actual parameters, the transient FEM simulation is performed to work out the field from the eddy current in the chamber conductor. Two cases are calculated, one with the vacuum chamber, the other without the chamber. The field difference between the two cases is the field produced only by the eddy current. The simulation result shows the angular distribution of the eddy current along the chamber wall is a cosine distribution with the peak current density $0.035A/mm^2$. According to (4), the peak density is $0.0347A/mm^2$, the agreement is very good.

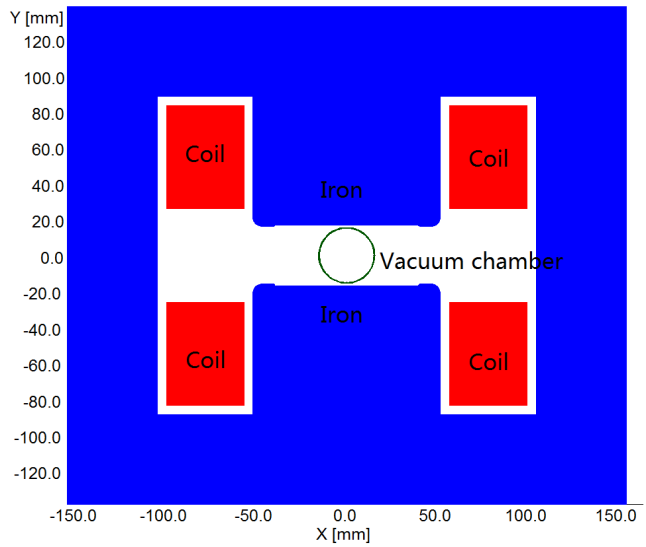

Fig.2. The cross section of HEPS dipole

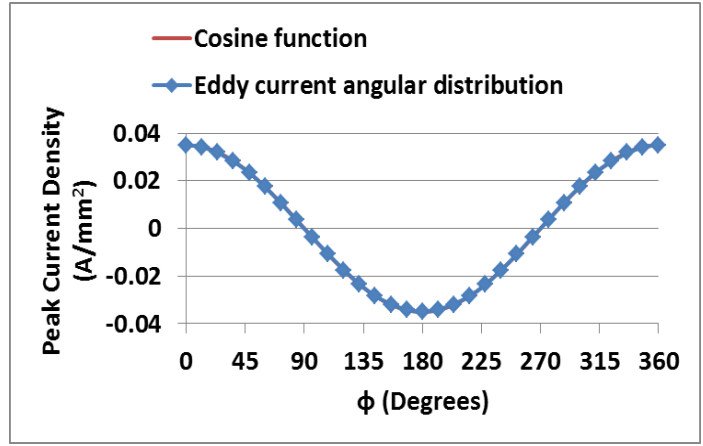

Fig.3. The angular distribution of the eddy current in vacuum chamber

According to (12), the sextupole component is $0.0166T/m^2$, which is in good agreement with $0.0154T/m^2$ by 2D FEM simulation. With the same parameters, the formula (1) gives a sextupole value of $0.12T/m^2$. The center fields $B_y(x=0, y=0)$ evaluated via (12) and FEM simulations are $0.33Gs$ and $0.28Gs$ respectively. Clearly, compared with (1), the expression (12) is more accurate and closer to the actual eddy current and field distributions.

## 5. Chromaticity

According to the discussions in Section 4, the real fields are the effective sextupole field superimposed on the nominal dipole fields, leading to changes in the chromaticity. If the eddy current is strong, it may cause the vertical chromaticity below zero, and induced beam head-tail instability and beam lost.

In the preliminary design, the chromaticity is corrected to (0.2, 0.2) without considering the eddy current effects. In ramping process, Based on (1), the horizontal and vertical chromaticity caused by the sextupoles in dipoles are $\xi_x$=0.58 and $\xi_y$=-1.14, the latter is smaller than zero, so we need compensate the chromaticity to a positive value in ramping process, but the big chromaticity maybe decrease the transverse mode coupling instability(TMCI) threshold [5] and dynamic aperture.

With the expression (12), the sextupole component due to the eddy current is smaller than that from (1) for one order of magnitude, so does the chromaticity. That means the chromaticity change in ramping can be neglected in HEPS booster case.

## 6. Conclusion

Using the complex contour integral, a closed-form magnetic field solution within the aperture of an induced circular vacuum chamber is obtained and testified by 2D FEM simulations. In Ref. [6], the magnetic field distribution in some neighborhood of $z=0$ is obtained by performing Taylor series expansion on (7), it is a little complicated to calculate coefficients of $z^n$, and then integrate them along the chamber wall. Expression (12) gives a global and unified description of the magnetic field within the aperture, and can be used for the multipole components estimations in preliminary accelerator design process.

**Acknowledgments**

This work is supported by the Foundation of Shanghai Ministry of Science and Technology, China, under Grant 90KF20171380. The author would like to acknowledge the supports and assistances from colleagues of IHEP.